\documentclass[runningheads]{svmult}

\usepackage{makeidx}   
\usepackage{graphicx}  
\usepackage{subeqnar}  
\usepackage{multicol}  
\usepackage{physmubb}  
\makeindex             

\newcommand{\Vec}[1]{\mbox{\boldmath$#1$}}
\begin{document}

\title*{Superconductivity from the repulsive electron interaction 
--- from 1D to 3D}

\toctitle{Superconductivity from the repulsive electron interaction  
--- from 1D to 3D}

\titlerunning{Superconductivity from the repulsive electron interaction}

\author{Hideo Aoki}

\authorrunning{Hideo Aoki}

\maketitle

An overview is given on how superconductivity 
with anisotropic pairing can be realised from 
repulsive electron-electron interaction.  
(i) We start from the physics in one dimension, where 
the Tomonaga-Luttinger theory predicts that, 
while there is no superconducting phase for the 
repulsive case for a single chain, the phase does 
exists in ladders with the number of legs equal to 
or greater than two, as shown both by analytically 
(renormalisation) and numerically (quantum Monte Carlo).  
(ii) We then show how this pairing has a natural extension 
to the two-dimensional 
case, where anisotropic (usually d) pairing superconductivity 
arises mediated by spin fluctuations (usually antiferromagnetic), 
as shown both by analytically 
(renormalisation) and numerically (quantum Monte Carlo).  
(iii) We finally discuss how the superconductivity from the 
electron repulsion can be ``optimised" (i.e., how $T_C$ 
can be raised) in 2D and 3D, where we propose that 
the anisotropic pairing is much favoured 
in systems having {\it disconnected Fermi surfaces} 
where $T_C$ can be almost an order of magnitude higher. 

\section{Introduction}

There is a growing realisation that the high-Tc 
superconductivity found in the cuprates in the 1980's 
has an electronic mechanism --- namely, anisotropic 
pairing from the repulsive electron-electron interaction.
Superconductivity from electron repulsion is conceptually 
interesting in its own right, and has indeed a long 
history of discussion.  In fact, in the field of 
electron gas, i.e., electron system with the Coulombic 
electron-electron interaction, Kohn and Luttinger\cite{KohnLutt} 
pointed out, as early as in the 1960's, that the 
electron gas should become superconducting with 
anisotropic pairing (having nonzero relative angular 
momenta) at sufficiently low temperatures in a 
perturbation theory.  
This becomes an exact statement for dilute enough electron 
gas, where p-wave (with the relative angular momentum = 1) 
should arise, as far as the static interaction is 
concerned\cite{Layzer,Takadaswave}.  

While these have to do with the long-range 
Coulomb interaction where the dominant fluctuation is 
charge fluctuation, the problem we would like to address 
here is the opposite limit of short-range repulsion, 
as appropriate for strongly-correlated systems such as 
transition metal oxides.  There, the dominant fluctuation is 
the spin fluctuation.  The most widely used model is the Hubbard 
model having the on-site repulsion, $U$.   
If the one-band Hubbard model, the simplest possible model 
for repulsively correlated electron systems, superconducts, 
the interest is not only generic but may be practical 
as well, which has indeed been a challenge in the 
physics of high $T_C$ superconductivity.  

To develop a theory for that, it is instructive to 
start with one-dimensional (1D) systems.  When the system is 
purely 1D, we have an exact effective theory, 
which is the Tomonaga-Luttinger theory and is 
exactly solvable in terms of the bosonisation and 
renormalisation.  So we start with this, where 
no superconducting phase is shown to exist 
when the interaction is repulsive.  
When there are more than one chains, 
which is called ladders, superconducting phase appears.  
If one closely looks at the pairing wavefunction, this 
is a pairing having opposite signs across 
two bands where the key process is the interband pair 
hopping.  

We then show that this physics has a very natural extension 
to two-dimensional(2D) systems.  There, anisotropic (usually d
having the relative angular momentum of 2) pairing 
superconductivity can arise.  
If one looks at the pairing wavefunction, this 
is a pairing having opposite signs across 
the key interband pair-hopping processes.  The 
key process is dictated by the peak in 
the spin structure (usually antiferromagnetic).  

We finally look at how this kind of anisotropic pairing 
superconductivity can be ``optimised", namely how we 
can make $T_C$ higher.   We first note that ``$T_C$ is 
very low in the electron mechanism" in that $T_C$ is 
usually two orders of magnitude lower than the electronic 
energy.  The main reason is the node in the gap function, 
which has to exist for the anisotropic pairing, intersects 
the Fermi surface.  So we can propose, and show, that systems that 
have disconnected Fermi surface has much higher $T_C$.

\section{1D --- Tomonaga-Luttinger theory and the physics of ladders}

\subsection{Tomonaga-Luttinger theory}

\begin{figure}
\begin{center}
\includegraphics[width=1.0\textwidth]{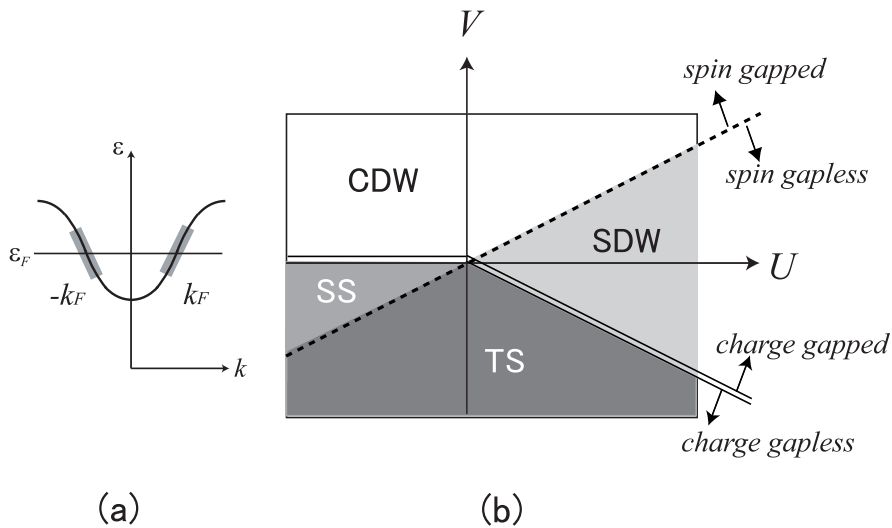}
\end{center}
\caption{(a) Tomonaga-Luttinger model, in which the 
low-energy excitations around the Fermi energy (shaded) at $k=\pm k_F$ 
are considered for 1D systems.
(b) Weak-coupling result for the phase diagram against 
the on-site repulsion, $U$, and the off-site repulsion, $V$, 
for the 1D extended Hubbard model at half filling. 
SS: spin-singlet superconductivity, TS: triplet superconductivity.
}
\label{TLmodel}
\end{figure}

It was Tomonaga who pioneered the many-body physics in 1D.  
In his 1950 paper\cite{Tomonaga} the essence of the whole idea is already 
there, although the theory is now often called Tomonaga-Luttinger.  
When the system is 1D, the Fermi 
energy, $E_F$, intersects the band at two points, 
left-moving branch ($L$) and the right-moving one ($R$; 
Fig.\ref{TLmodel}(a)).  
The dispersion around these points 
may be approximated as linear functions of the 
wavenumber, $k$.   When we do this, every electron-hole excitation 
across $E_F$ becomes a creation operator of a sound wave 
(which is a boson).  

As for the electron-electron interaction, 
the matrix elements may be classified into four categories: 
(i) backward scattering, where one electron at $R$ jumps to 
$L$ while another from $L$ to $R$, 
(ii) forward scattering, where  $R$ jumps to 
$R$, $L$ jumps to $L$, 
(iii) umklapp scattering, where ($[R,R]$ jumps to $[L,L]$ or vice versa), 
(iv) forward scattering within each branch ($[R,R]$ to $[R,R]$ or 
$[L,L]$ to $[L,L]$).  
Tomonaga-Luttinger theory
~\cite{TomLutreview,weak1,weak2,weak3,weak4,weak5}
is a weak-coupling theory (i.e., theory for the case 
when the electron-electron interaction is weak enough), where 
only question low-energy processes.  For that we can integrate out 
the higher-energy processes in the perturbational renormalisation-group 
sense.  We can then look at the flow of the renormalisation 
equation, and its end point called the fixed point.  
To discuss the nature of the fixed-point Hamiltonian, it is 
convenient to bosonise (i.e., to write everything in terms of 
boson operators).  The final result for the effective Hamiltonian 
is written in terms of two boson fields, spin phase ($\phi$) and 
charge phase ($\theta$), whose stiffness (coefficients of 
$(\partial \phi)^2, (\partial \theta)^2$) is given 
in terms of only two quantities, $K_{\sigma}, K_{\rho}$, 
which determine everything, including whether the 
ground state is superconducting.   
To be more precise, in 1D even a ``long-range" 
order can only have a two-point correlation that decays 
with a power law ($\propto 1/r^{\alpha}$) where the exponent 
$\alpha$ is dictated, for each of the 
order parameters considered, by $K_{\sigma}, K_{\rho}$. 

For every Hamiltonian originally given, we can calculate 
the four scattering parameters, and then renormalise them.  
If we look at the phase diagram (Fig.\ref{TLmodel}(b)) 
for the extended Hubbard model 
(where we have an off-site interaction, $V$, on top of 
the on-site one, $U$), 
there is no superconducting phase when all the interactions 
$U$ and $V$ are repulsive ($>0$).


Incidentally, there is no magnetism, either, for a single chain.  
This is due to the well-known Lieb-Mattis theorem, which dictates 
that electrons in 1D are entirely non-ferromagnetic.  
The proof makes use of the fermion statistics of electrons, where 
a key factor is no two electrons can pass each other in 1D, or, 
in the words in Mattis's textbook\cite{Mattis}, ``neighbours 
remain neighbours till death did them part".  

\subsection{Pairing in ladders, or $1+1\neq 2$}

\begin{figure}
\begin{center}
\includegraphics[width=1.0\textwidth]{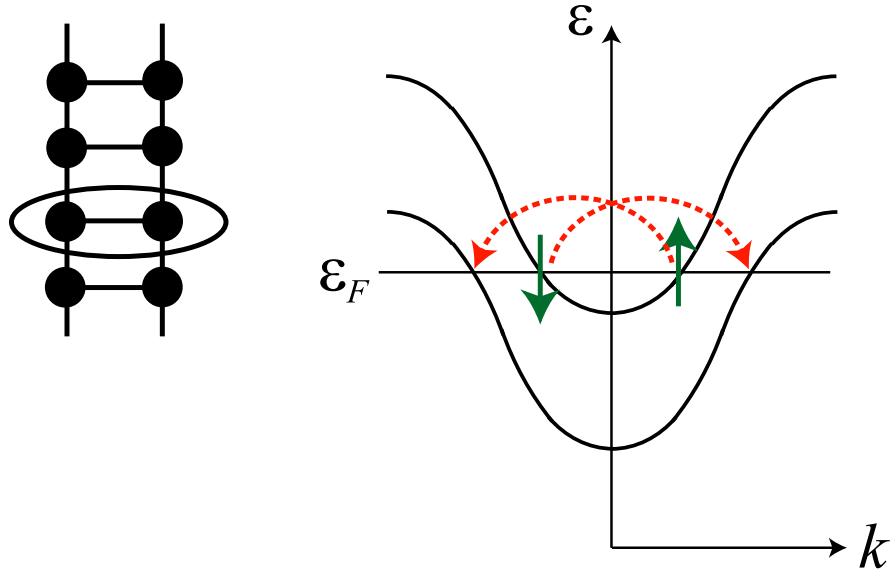}
\end{center}
\caption{Multi-band 1D model (right panel) 
for the Hubbard model on a ladder (left).  
An oval on the ladder represents the inter-chain pairing.
}
\label{ladderdispersion}
\end{figure}

When there are more than one chain with inter-chain 
hopping and/or interaction, the physics can be, and is 
indeed, entirely different.  
The model, then, becomes multi-band (i.e., $n$-band 
system for $n$-leg ladder).  The Fermi energy 
can intersect the dispersion at $2n$ points (Fig.\ref{ladderdispersion}), so 
the model is what can be called {\it multiband Tomonaga-Luttinger model}.  
The multiband Tomonaga-Luttinger model has been studied 
in various context, including 
the excitonic phase in electron-hole systems\cite{nagaosaogawa}, 
transport properties\cite{spinpolTL} and interband 
excitations in quantum wires as 
detected by Raman spectroscopy by Sassetti et al.\cite{sassetti}  

In the context of the high $T_C$, 
the idea of superconductivity in multi-chain (or ``ladder") 
systems was kicked off theoretically in 1986, 
when Schulz~\cite{SchulzAF} proposed a possible relation 
between ladders at half-filling and Haldane's conjecture 
for spin chains.   
He made the following reasoning: 
If we consider repulsively interacting electrons 
on a ladder, the undoped system will be 
a Mott insulator, so that we may consider the system as 
an $S=1/2$ antiferromagnetic (AF) Heisenberg magnet on a ladder 
for large Hubbard repulsion $U$. 
Schulz's analysis~\cite{SchulzAF} is that 
an AF $S=n/2$ single chain, which is exactly the
Haldane's system~\cite{Haldane,Nishiyama}, is similar to 
an $S=1/2$ AF ladder with $n$-legs. 
For the spin chains, Haldane~\cite{Haldane} has conjectured that 
the spin excitation should be gapless for half-odd-integer spins 
($n$: odd) or gapful for integer spins ($n$: even).  
If the situation is similar 
in ladders, a ladder having an even number of legs 
will have a spin gap, associated with 
a `spin-liquid' ground state where the quantum 
fluctuation is so large that the AF correlation 
decays exponentially.  

Dagotto {\it et al.}\cite{Dagotto1} and by Rice {\it et al.}\cite{Rice}, 
then suggested the possibility of
superconductivity associated with the spin gap.  
The presence of a spin gap, i.e., a gap in the spin excitation 
which is indicative of a quantum spin liquid, in the two-leg ladder 
(or, more generally, in even legs) is a good news for 
superconductivity, since an idea proposed by Anderson~\cite{Anderson}
in the context of the high-$T_C$ superconductivity 
suggests that a way to obtain superconductivity is to 
carrier-dope spin-gapped systems. 
The superconductivity in even-leg ladders 
is in accord with this.
Subsequently superconductivity 
has been reported for a cuprate with a ladder structure\cite{Uehara}, 
although it later turned out that this material has 
a rather strong two-dimensionality that may dominate 
the superconductivity.

For doped systems, the conjecture for superconductivity~\cite{Rice} 
is partly based on an exact diagonalisation study
for finite $t$-$J$ model on a two-leg ladder.~\cite{Dagotto} 
This was then followed by analytical~\cite{Sigrist}
and numerical~\cite{Poilblanc3,Tsunetsugu,Hayward,Sano}
works on the doped $t$-$J$ ladder, for which 
the region for the dominant pairing
correlation appears at lower side of the exchange coupling $J$ 
than in the case of a single chain. 

On the other hand the Hubbard model on a ladder is of 
general interest.\cite{KAmultiTL}
Although the Hubbard crosses over to the 
$t$-$J$ model for $U\rightarrow \infty$, 
we have only an infinitesimal $J$ there, so 
the result for $t$-$J$ model does not directly answers this.  
Since there is no exact solution for the Hubbard ladder, 
we can proceed in two ways: 
for small $U$ we can adopt an analytic method, which 
is the weak-coupling renormalisation-group theory,
where the band structure around the Fermi points is linearised
in the continuum limit 
to treat the interaction with a perturbative renormalisation group. 
The weak-coupling theory with the bosonisation and
renormalisation-group techniques has been 
applied to the two-leg Hubbard ladder.
~\cite{Finkelstein,Balents,Fabrizio,Fab,Nagaosa,Schulz2} 

The Hamiltonian of the two-leg Hubbard ladder 
is given in standard notations as 
\begin{eqnarray}
{\cal H}&=&-t\sum_{\alpha\langle i\rangle \sigma}
(c_{i\sigma}^{\alpha \dagger} c^{\alpha}_{i+1\sigma}+{\rm h.c.})
-t_{\perp}\sum_{i \sigma}
(c_{i\sigma}^{1 \dagger} c^{2}_{i\sigma}+{\rm h.c.})\nonumber \\&&
+U\sum_{\alpha i} n^{\alpha}_{i\uparrow}n^{\alpha}_{i\downarrow},
\end{eqnarray}
where $\alpha(=1,2)$ specifies the chains, 
or in the momentum space as
\begin{eqnarray}
{\cal H}&=&-2t\sum_{\mu k\sigma}\cos(k) c^{\mu \dagger}_{k\sigma}
c^{\mu}_{k\sigma} -2t_{\perp}\sum_{k\sigma}c^{0 \dagger}_{k\sigma}
c^{0}_{k\sigma}\nonumber\\&&
+U\sum ({\rm interaction\:\:of\:\:the\:\:form}\:\: c^\dagger c^\dagger c c),
\end{eqnarray}
where $\mu$ specifies the bonding ($\mu=0$) and anti-bonding ($\mu=\pi$) bands, 
so labelled since $k_y=0,\pi$, respectively. 

The part of the Hamiltonian, $H_{\rm d}$, that can be 
diagonalised in the bosonisation includes only 
intra- and inter-band forward-scattering processes 
arising from the intrachain forward-scattering terms. 
We can then define bosonic operators as in the 
single-chain case. 
If we introduce the phase variables as in the single-chain case, 
$H_{\rm d}$, written in terms of them, is 
separated into the spin-part $H_{\rm spin}$
and the charge-part $H_{\rm charge}$.
While $H_{\rm spin}$ is already diagonalised, 
$H_{\rm charge}$ can be made so 
with a linear transformation, and the diagonalised 
$H_{\rm charge}$ is written in terms of 
the correlation exponent $K_{\rho i}, i=1,2$.
So we end up with the total Hamiltonian that reads
\begin{equation}
H = {\rm kinetic energy} + H_{{\rm d}}+\mbox{pair-hopping terms}.
\end{equation}
Here, the pair-hopping (or pair-scattering) term represents 
those part of the interaction Hamiltonian, 
in which a pair of electrons is scattered via the interaction 
to another pair of electrons.

At half-filling, the system reduces to a spin-liquid 
insulator having both charge and spin gaps~\cite{Balents}. 
When carriers are doped to the two-leg Hubbard ladder, 
on the other hand, the relevant scattering processes 
at the fixed point in the renormalisation-group flow 
are the pair hopping across 
the bonding and anti-bonding bands (Fig.\ref{ladderdispersion}), 
$c_{\uparrow}^{\pi \dagger} c_{\downarrow}^{\pi \dagger} c_{\downarrow}^0 
c_{\uparrow}^0+{\rm h.c.}$ in $k$ space, 
and the backward-scattering process within each band. 
The importance of the pair-hopping across the two bands 
for the dominance of pairing correlation in the two-leg Hubbard ladder
is reminiscent of the Suhl-Kondo mechanism, which 
was proposed back in the 1950's for superconductivity
in a quite different context of the s-d model for 
the transition metals.~\cite{Suhl,Kondo} 

The renormalisation results in a formation of gaps 
in both of the two spin modes and a gap in one of the charge modes. 
This leaves one charge mode massless, where the mode is characterised by 
a critical exponent $K_\rho$.  Then the correlation of the  
intraband singlet pairing,
\begin{equation}
\sum_\sigma \sigma(c_{k\sigma}^0 c_{-k,-\sigma}^0
-c^{\pi}_{k\sigma}c^{\pi}_{-k,-\sigma}),
\label{2legpair}
\end{equation}
decays like $1/r^{1/(2K_\rho)}$, where $K_\rho$ 
should be close to unity in the weak-coupling regime. So 
this should be the dominant phase, which is, 
expressed in real space as 
$c^1_{i\sigma} c^2_{i,-\sigma}-c^1_{i,-\sigma} c^2_{i\sigma}$, 
an {\it interchain} singlet pairing. 

\subsection{How to detect pairing in quantum Monte Carlo studies?}

The perturbational renormalisation group 
is in principle guaranteed to be valid only for sufficiently 
small interaction strengths ($U\ll t$), 
so that its validity for finite $U (\sim t$) 
has to be checked.  
To be more precise, the renormalisation approach can tell 
whether the interaction flows into weak coupling 
(with the relevant mode gapless) or into 
strong-coupling regime (gapful) for small enough 
interactions, but the framework itself (i.e., the 
perturbational expansion) might fail for stronger interactions.  

This is where numerical studies come in.  
Numerical calculations for finite $U$ have been performed 
with the exact diagonalisation, DMRG 
or quantum Monte Carlo (QMC) methods,~\cite{Hirsch,MC1,MC2,MC3,MC4,MC5} 
but in an earlier stage the results are scattered, where 
some of the results seemed inconsistent with the weak-coupling prediction:  
a DMRG study by Noack {\it et al.} for the doped Hubbard ladder 
shows the enhancement of the pairing correlation
over the $U=0$ result strongly depends on 
the inter-chain hopping, $t_{\perp}$\cite{Noack1,Noack2}. 
Quantum Monte Carlo (QMC) results also exhibit an absence\cite{Asai} 
or presence\cite{Yamaji} of the enhancement depending 
on the hopping parameters and/or band filling.

Recently, however, a QMC study by Kuroki et al~\cite{Kuroki}  
has resolved the puzzle, and has 
clearly detected an enhanced pairing correlation.  
A key factor found there in detecting  
superconductivity in any numerical calculation, which also 
resolves the origin of the former discrepancies is: 
we have to question a very 
tiny energy scale ($\ll$ starting electronic 
energy scale, $t,U$) in detecting the 
pairing.  This immediately implies that 
the discreteness of energy levels in finite systems 
examined in QMC studies enormously affects 
the pairing correlation --- If the level separation 
is greater than the energy scale we want to look at, 
any feature in the 
correlation function will be easily washed out.   
This can be circumvented 
if we tune the parameters so as to make the 
separation between the levels just below and above 
$E_F$ tiny (i.e., to make the LUMO-HOMO nearly degenerate 
in the quantum chemical language), 
which should be a reasonable way to approach the 
bulk limit where the levels are dense.  
The importance of small offsets between the highest occupied and 
lowest unoccupied levels has also been stressed by
Yamaji {\it et al.} for small systems.~\cite{Yamaji}

\begin{figure}
\begin{center}
\includegraphics[width=0.75\textwidth]{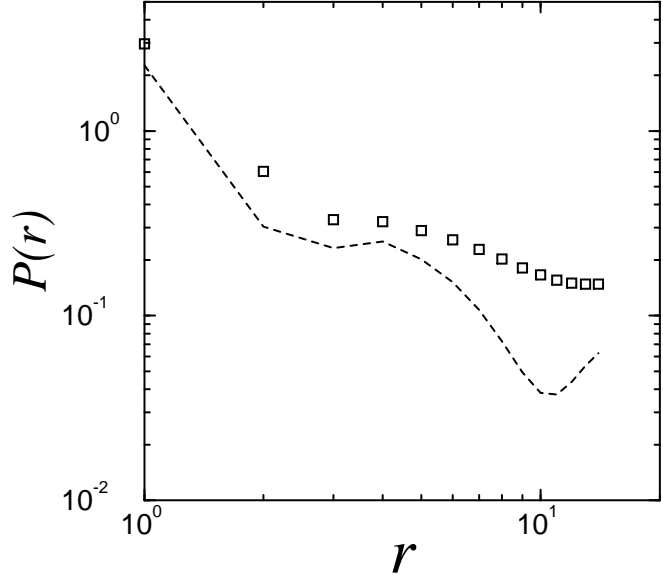}
\end{center}
\caption{
The pairing correlation function plotted 
against the real space distance $r$ in a 30-rung 
Hubbard ladder having 56 electrons 
for $U=1$ with $t_{\perp}=1.975$ (square).\cite{Kuroki} 
The dashed line is the non-interacting result.
}
\label{ladderQMC}
\end{figure}

So we have applied the (projector) Monte Carlo method\cite{pmc}
to look into the ground state correlation function 
$P(r)\equiv \langle O_{i+r}^{\dagger} O_{i}\rangle$
of this pairing for finite values of $U (\sim t)$.  
We show in Fig.\ref{ladderQMC} the result for $P(r)$
for $t_{\perp}=0.98$, $U=1$ and 
the band filling $n=0.867=52$ electrons/ (30 rungs $\times$ 2 sites).  
The $U=0$ result (dashed line) 
for these two values of $t_{\perp}$ are identical 
because the Fermi sea remains unchanged.
If we turn on $U$, we can see that a large
enhancement over the $U=0$ result emerges at large distances. 

We have deliberately chosen the value of $t_{\perp}=0.98$ 
to make the one-electron energy levels of the bonding and  
anti-bonding bands lie close to each other 
around the Fermi level within $0.004t$.  
This is much smaller than the energy scale we question 
(which is the spin gap in the present 1D case).  
In fact, a 5\% change in $t_{\perp}=0.98\rightarrow 1.03$, 
for which the LUMO-HOMO separation blows up to $\sim 0.1t$, 
washes out the enhancement in the correlation function. 
In the latter case 
the renormalisation of higher energy modes has to stop 
at this energy scale, so that the interband pair hopping process 
will not be renormalised into a strong coupling, while  
in the weak-coupling theory the renormalisation 
all the way down to the Fermi level is assumed. 

While it is difficult to determine the decay exponent
of the pairing correlation $P(r)$, we can fit the 
data by assuming a trial function expected from the weak-coupling theory, 
$P(r) \propto c/\sqrt{r} + (1-c)/r^{2}
+[\cos(2k_F^0 r)+\cos(2k_F^{\pi} r)]/r^{2}$
where 
the overall decay at large distances is assumed to be $\propto 1/\sqrt{r}$ 
as dictated in the weak-coupling theory.  This form reproduces the result 
surprisingly accurately. 


\section{Three-leg ladder and 1D-2D crossover}

Now, the physics of ladders can provide 
quite an instructive line of approach for understanding 
the physics in two-dimensional systems via the crossover 
from 1D to 2D (two-dimensions).  So let us first look at the three-leg ladder.

\subsection{Three-leg ladder}

If we return to ladders, one can naively expect that 
ladders with odd-number (e.g., 3) of legs will have no spin gap, 
which would then signify an absence of 
dominating pairing correlation 
(`even-odd conjecture for superconductivity').
As far as the spin gap in undoped ladders is concerned, 
experiments on 
a class of cuprates, Sr$_{n-1}$Cu$_{n}$O$_{2n-1}$ 
having $n$-leg ladders, have supported the conjecture.
\cite{Azuma,Ishida1,Kojima,Hiroi2,Mayaffre} 
So it was believed that odd-numbered legs 
only have the usual $2k_F$ spin-density wave (SDW) rather than 
superconductivity. 

Kimura et al\cite{Takashi1,Takashi2},however, showed that that is 
too simplistic a view, and that, while the 
even-odd conjecture for the spin gap is certainly correct, 
an odd-number of legs does indeed superconduct by exploiting the 
spin-gapped mode.  In that work 
the pairing correlation in the three-leg 
Hubbard ladder has been examined\cite{tJcom}. 

\begin{figure}
\begin{center}
\includegraphics[width=0.8\textwidth]{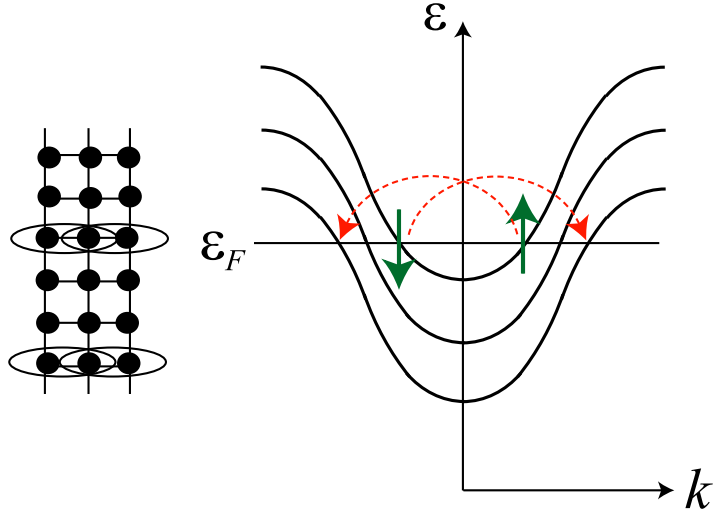}
\end{center}
\caption{
Three-leg ladder (inset) and its dispersion.  
The dashed arrows represent the pair-hopping process, 
and ovals represent the inter-chain pairing.
}
\label{threeleg}
\end{figure}

We start with the weak-coupling theory for correlation functions 
for the three-leg Hubbard ladder (Fig.\ref{threeleg}). 
Arrigoni has looked into a three-leg ladder 
with weak Hubbard-type interactions with 
the perturbational renormalisation-group technique 
to conclude that gapless and gapful spin excitations 
coexist in three legs.~\cite{Arrigoni} 

He has actually enumerated  the numbers of gapless 
charge and spin modes on the phase diagram spanned by 
the doping level and the interchain hopping, $t_{\perp}$.
He found that, at half-filling, one gapless spin mode exists.  
For general band filling, one gapless spin mode remains 
in the region where the Fermi level intersects all the 
three bands in the noninteracting case. From this, Arrigoni argues that 
the $2k_F$ SDW correlation should decay as a power law
as expected from experiments. 
Arrigoni's result indicates that two gapful spin modes 
exist in addition. 
The charge modes, on the other hand, consists of two 
gapless modes and one gapful mode.  


The question we address then is what happens 
when gapless and gapful spin modes {\it coexist}.  This is an 
intriguing problem, since 
it may well be possible that the presence of gap(s) in 
{\it some} out of multiple spin modes may be sufficient 
for the dominance of a pairing correlation.
Schulz~\cite{Schulz3} has independently
shown similar results for a subdominant $2k_F$ SDW and 
the interchain pairing correlations. 

Physically, the picture that emerges as we shall describe below, 
is that the two spin gaps, which are relevant to the pairing, arise 
as an effect of the pair-hopping process  
that is the many-body matrix element transferring 
two electrons simultaneously across the outermost bands 
(i.e., the top and bottom bands for a three-leg ladder) (Fig.\ref{threeleg}). 
In this sense the mechanism is 
reminiscent of the situation in the two-leg case or 
the Suhl-Kondo mechanism.~\cite{Suhl,Kondo}

The correlation functions can be calculated with 
the bosonisation method~\cite{TomLutreview} 
for the three-leg Hubbard model.

We can define three bosonic operators, 
where we diagonalise $H_{\rm charge}$ in terms of 
the three correlation exponents $K_{\rho i}, i=1,2,3$.
As Arrigoni pointed out~\cite{Arrigoni} 
the pair-hopping processes across the top 
and bottom bands become relevant as the renormalisation is performed. 
In order to actually calculate the correlation functions, 
we have to express the relevant scattering processes in terms of the 
phase variables. The fixed-point Hamiltonian density, 
$H^{*}$, takes the form, in terms of the phase variables, 
\begin{eqnarray}
H^{*} \propto &-&g_{\rm back}(1){\rm cos}[2\phi^{1+}(x)]
            -g_{\rm back}(3){\rm cos}[2\phi^{3+}(x)]\nonumber \\
         &+&2g_{\rm pair hopping}(1,3) {\rm cos}[\sqrt{2}\chi^{1-}(x)]
            {\rm sin}\phi^{1+}(x){\rm sin}\phi^{3+}(x),
\end{eqnarray}
where $g_{\rm back}(1), g_{\rm back}(3)$ are negative large quantities, 
and $g_{\rm ph}(1,3)$ is a positive large quantity. 
This indicates the following.  
Two spin phases, $\phi^{1+}$, $\phi^{3+}$, 
become long-range ordered and fixed, 
respectively, while $\phi^{2+}$ is not fixed to give a gapless spin mode.  
Similarly, the difference in the charge phases, $\theta^i$, 
for the outermost bands,
\begin{equation}
\chi^{1-} \equiv \frac{1}{\sqrt{2}}(\theta^{1-} - \theta^{3-}), 
\end{equation}
is ordered and fixed, 
and the charge gap opens for this particular mode.

Now we can calculate the correlation functions,
since the gapless fields have 
already been diagonalised, while the remaining gapful fields have 
the respective expectation values. 


Among various order parameters, the dominant one (with the 
longest tail in the correlation) is the 
singlet pairing across the central and edge chains (Fig.\ref{threeleg}), 
which is, in the band picture, 
\begin{eqnarray}
O_{d} \sim \sum_{\sigma}\sigma(\psi^{1+}_{\sigma}\psi^{1-}_{-\sigma}
-\psi^{3+}_{\sigma}\psi^{3-}_{-\sigma}). 
\end{eqnarray} 
We call this ``d" for the following reason.  
Since we have taken a continuum limit along the chain, 
it is not straightforward to 
name the symmetry of a pairing.  However, 
we could call the above pair as d-wave-like 
in that the pairing, in addition to being off-site on the rung, 
is a linear combination of 
a bonding band and an anti-bonding band with opposite signs.  
Since the relevant pair-hopping is across these bands, 
we can say that there is a node in the pair wavefunction 
along the line that bisects the relevant pair-hopping.  

Calculation of the  correlation function of $O_{d}$ gives
\begin{equation}
\langle O_{d}(x)^{\dagger}O_{d}(0)\rangle 
\sim 
x^{-\frac{1}{3}(\frac{1}{K_{\rho2}^*}+\frac{1}{2K_{\rho3}^*})},
\end{equation}
which is the dominant ordering.  
In the weak-interaction limit ($U\rightarrow +0$), 
where all the $K^*$'s tend to unity, the ``d"-pairing 
correlation decays as slowly as $x^{-1/2}$, 
while the other correlations decay like $x^{-2}$.
We can see that the interchain pairing {\it exploits} 
the charge gap and the spin gaps to reduce the exponent of the correlation
function, in contrast to the intrachain pairing.  
Namely, we would have ($-1$) in the exponent if the spin were 
gapless.  This alone would only result in a $1/r$ decay, but 
the charge mode $\chi_{1-}$ is further locked, 
which further reduces the exponent (down to $1/\sqrt{r}$ 
in the limit $U\rightarrow 0$).  

Now, how the pairing 
correlation in the three-leg Hubbard ladder looks like when $U$ 
is finite? 
Our QMC result for the three-leg Hubbard ladder
exhibits an enhancement of 
the pairing correlation even for finite coupling constants, 
$U/t=1\sim 2$.~\cite{Takashi2} 
As in the two-leg case with a finite $U$, 
we have taken care that 
levels below and above the Fermi level are close.  



\subsection{1D-2D crossover}

We have seen that the weak-coupling theory (perturbational renormalisation +
bosonisation) predicts that 
the interband pair hopping 
between the innermost and outermost Fermi points $(k_x, k_y) \approx 
(\pm k_F^0,0), (\pm k_F^{\pi},\pi)$ 
becomes relevant (i.e., increases with the renormalisation). 
This concomitantly makes the two-point 
correlation of the interchain singlet 
decay slowly with distance. 
In $k$-space the dominant component of this pair reads
\begin{equation}
\sum_{\sigma}\sigma\
(c_{k_F^0,\sigma}^0 c_{-k_F^0,-\sigma}^0-
c_{k_F^\pi,\sigma}^\pi c_{-k_F^\pi,-\sigma}^\pi).
\label{eqn1}
\end{equation}

\begin{figure}
\begin{center}
\includegraphics[width=1.0\textwidth]{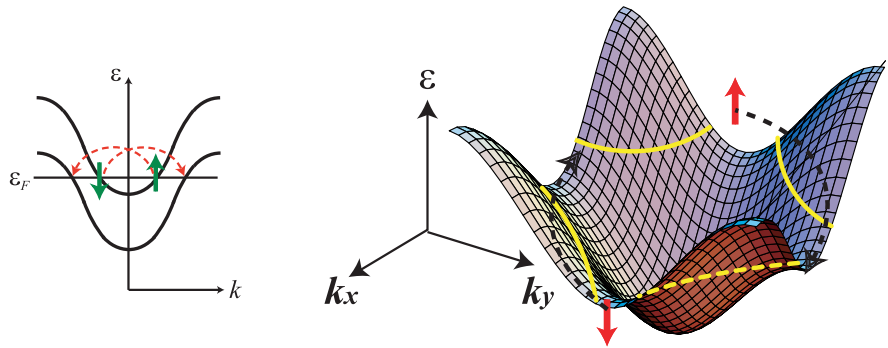}
\end{center}
\caption{
Band dispersion of the square lattice, on which 
the Fermi surface for $E_F$ close to half filling 
and a typical pair-hopping process (dashed arrows) 
are shown.  Similar plot for a ladder is attached for comparison.
}
\label{sqdisppairhop}
\end{figure}

Now, when $E_F$ intersects 
the outermost-band top and the innermost-band bottom with 
$k_F^0\simeq \pi, k_F^\pi\simeq 0$ (Fig.\ref{sqdisppairhop} 
left), 
{\it intrachain nearest-neighbor} singlet pair also has 
a dominant Fourier-component equal to eqn.(\ref{eqn1}) 
with a phase shift $\pi$ relative to the interchain
pairing. Thus, a linear combination 
which amounts to the d$_{x^2-y^2}$ pairing 
should become dominant.  We shall see this is exactly 
what happens in the 2D squre lattice around the half filling, 
Fig.\ref{sqdisppairhop} right.

\section{Superconductivity from the repulsive interaction in 2D}

As we have seen, the two-leg and three-leg 
Hubbard ladders do superconduct, so what will happen 
if we consider $n$-leg ladders for $n=3,4,...,\infty$ 
to reach the 2D square lattice.  
This view enables us to have a fresh look at the 2D Hubbard model.  

So we move on to the repulsive Hubbard model on the square lattice.  
The seminal notion that the high-$T_C$ conductivity in cuprates should 
be related the strong electron correlation was first put forward 
by Anderson.\cite{Andersonreview}  
There the superconductivity is expected 
to arise from the pairing interaction mediated by 
spin fluctuations (usually antiferromagnetic).  
A phenomenology along this line such as the self-consistent renormalisation
\cite{Moriya1,Moriya2,Moriya3,Pines1} has 
succeeded in reproducing anisotropic $d$-wave superconductivity.  
Microscopically, the repulsive Hubbard model, a simplest 
possible model for correlated electrons, 
should capture the physics in cuprates.\cite{dpKA}  
Some analytical calculations have suggested the occurrence
of d$_{x^2-y^2}$-wave superconductivity in the 2D Hubbard model. 
\cite{Bickers,Dzyalo,Schulz,ScalapinoPhysRep,Alvarez}
In particular, fluctuation exchange approximation (FLEX), 
developed by Bickers {\it et al.}\cite{FLEX}, 
has also been applied to 
the Hubbard model on the square lattice
\cite{Dahm,Deisz} to show the occurrence of the superconductivity.  

Numerical calculations have also been performed extensively.  
Finite binding energy\cite{DMST,FOP}
and pairing interaction vertex\cite{White,Husslein,Zhang} were
found in those calculations. 
Variational Monte Carlo calculations show that a superconducting 
order lowers the variational energy.\cite{Giamarchi,YamajiVMC}
Nevertheless, there had been a reservation against the occurrence of 
superconductivity in the Hubbard model because 
the pairing correlation functions
do not show any symptom of long-range 
behavior in some of the works.\cite{Zhang,Furukawa,Moreo}  

Again, Kuroki et al\cite{Kuroki} showed for the first 
time that QMC does indeed exhibit symptoms of 
superconductivity 
if we take proper care of a small energy scale involved, 
i.e., the d-wave pairing correlation becomes 
long-tailed when the Fermi level lies between 
a narrowly separated levels residing on the k-points 
across which the dominant pair hopping occurs.  
An enhancement of the 
pairing correlation has in fact 
been found by exact diagonalisation\cite{Yamaji} 
and by density matrix renormalisation group\cite{Noack2}
when $E_F$ lies close to the $k$-points 
$(0,\pi)$ and $(\pi,0)$.
Although the d$_{x^2-y^2}$-like nature of the pairing was 
suggested,\cite{Noack2} d$_{x^2-y^2}$ pairing correlation 
itself has not been calculated.  
So, in our quest for 2D, 
we first calculate the correlation function with QMC.
Here we employ the ground-state, 
canonical-ensemble QMC,\cite{stab} where we take the free 
Fermi sea as the trial state.

\subsection{Anisotropic pairing in 2D}

First, let us look at why the attractive interaction is 
by no means a necessary condition for superconductivity, 
which can quite generally arise from repulsive electron-electron 
interactions, which seems to be still not realised well enough.   
If we look at the BCS gap equation, we can immediately 
see that superconductivity can readily arise 
from repulsive interactions.  The gap equation reads
\begin{eqnarray}
1&=&-V_\phi\sum_{\Vec{k}'}\frac{1}{2\varepsilon(\Vec{k}')}
{\rm tanh}\left[\frac{1}{2}\beta \varepsilon(\Vec{k}')\right],
\nonumber 
\\
&&V_\phi=\frac{\langle V(\Vec{k,k}')\Delta(\Vec{k})\Delta(\Vec{k}') 
\rangle_{\rm FS}}
{\langle\Delta^2(\Vec{k})\rangle_{\rm FS}},
\end{eqnarray}
where $\varepsilon(\Vec{k})$ is the band energy measured from 
the chemical potential, $\beta =1/(k_BT_C)$, 
$V(\Vec{k,k}')$ the pair-hopping matrix element, 
$\Delta(\Vec{k})$ the BCS gap function, and 
$\langle ... \rangle_{\rm FS}$ is the average over the 
Fermi surface.  
So, if $\Delta(\Vec{k})$ has nodes across $\Vec{k} \leftrightarrow \Vec{k}'$ 
(i.e., changes sign before and after the pair hopping), 
the originally repulsive $V>0$ acts effectively as an attraction, 
\[
\langle V(\Vec{k,k}')\Delta(\Vec{k})\Delta(\Vec{k}') 
\rangle_{\rm FS} < 0.  
\]
This most typically happens for the $d_{x^2-y^2}$ pairing with 
$\Delta(\Vec{k}) \propto {\rm cos} (k_x)-{\rm cos} (k_y)$ when 
the dominant pair hopping occurs across 
$\Vec{k}\sim(0,\pi) \leftrightarrow \Vec{k}'\sim (\pi,0)$.  

When the spin fluctuation is antiferromagnetic, 
most typically in bipartite lattices such as a square lattice, 
the importance of the interactions around 
$(0,\pi)$ and $(\pi,0)$
in the 2D Hubbard model has been suggested by 
various authors.
\cite{Dzyalo,Schulz,ScalapinoPhysRep,Alvarez,Husslein,YamajiVMC,LeeRead,Newns,Mark,Gonzalez}

Group theoretically, the square lattice has a tetragonal 
symmetry, so that everything, including the gap function, 
should be an irreducible representation of the tetragonal 
group.  The $d_{x^2-y^2}$ pairing indeed belongs to B$_{1g}$ 
representation of this group.

\subsection{Quantum Monte Carlo study for the 2D Hubbard model}

In the context of our QMC study for the 2D Hubbard model, 
we have to take finite systems that 
have the $k$-points around $(0,\pi)$ and $(\pi,0)$ close in
energy. Nameley, our expectation from the study on ladders is that the 
pair hopping processes across around 
$(0,\pi)$ and $(\pi,0)$ may result in d$_{x^2-y^2}$ pairing, 
$\sum_{\Vec{k}} [{\rm cos}(k_x)-{\rm cos}(k_y)] 
c_{\Vec{k}\uparrow}c_{-\Vec{k}\downarrow}$ in 2D,
but an enhanced pairing correlation
should be detected only when the level offset between
the discrete levels around those points is small.

\begin{figure}
\begin{center}
\includegraphics[width=1.0\textwidth]{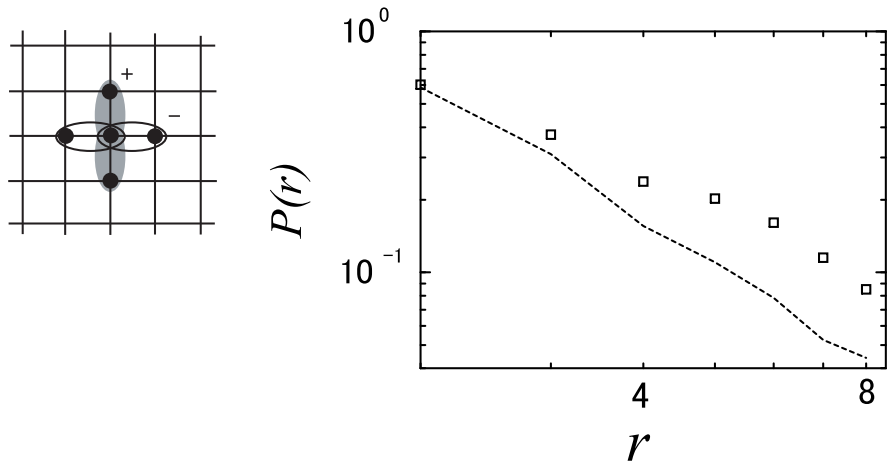}
\end{center}
\caption{
QMC result for the d$_{x^2-y^2}$ pairing correlation
for a $10\times 10$ square lattice with 78 electrons 
for $U=1$ and $t_y=0.999$ (square).\cite{Kuroki} 
The dashed line represents the noninteracting case.
Inset depicts the square lattice with the ovals 
representing a d-wave pair.
}
\label{hubbardQMC}
\end{figure}

We take 78 electrons in $10\times 10$ sites ($n=0.78$)
with $t_y=0.999$ with periodic boundary condition in both directions.  
We have taken $t_y=0.999$, because the number of electrons considered 
here would have an open shell (with a degeneracy in the free-electron 
Fermi sea) for $t_y=1$, which will destabilise QMC convergence. 
Taking $t_y=0.999$ lifts the degeneracy 
to give a tiny ($<0.01$) but finite $\Delta\varepsilon^0$.
In Fig.\ref{hubbardQMC} we plot the d$_{x^2-y^2}$ pairing correlation, 
defined as
\begin{eqnarray}
P(r)&=&\sum_{|\Delta x|+|\Delta y|=r} 
\langle O^{\dagger}(x+\Delta x,y+\Delta y) O (x,y)\rangle, \nonumber \\
O(x,y)&=&\sum_{\delta=\pm 1,\sigma}
\sigma(c_{x,y,\sigma} c_{x+\delta,y,-\sigma}
-c_{x,y,\sigma} c_{x,y+\delta,-\sigma}),
\end{eqnarray}
where the correlation for $U=1$ is clearly seen to be enhanced 
over that for $U=0$ especially at large distances. 

We can readily show that when the level spacing becomes too 
large (e.g., $\Delta\varepsilon^0 \sim 0.1$), 
the enhancement is washed out. 
In the present choice the energy levels around 
$(0,\pi), (\pi,0)$ are close 
($< 0.01t)$, while the other 
levels lie more than $\sim 0.1t$ away from $E_F^0$.  
One might thus raise a criticism that the scattering processes involving 
the states away from $E_F^0$ are unduly neglected.  
We can however show (not displayed here) 
that when other levels exist around $E_F$ an enhanced 
d$_{x^2-y^2}$ correlation is obtained as well.

\begin{figure}
\begin{center}
\includegraphics[width=0.8\textwidth]{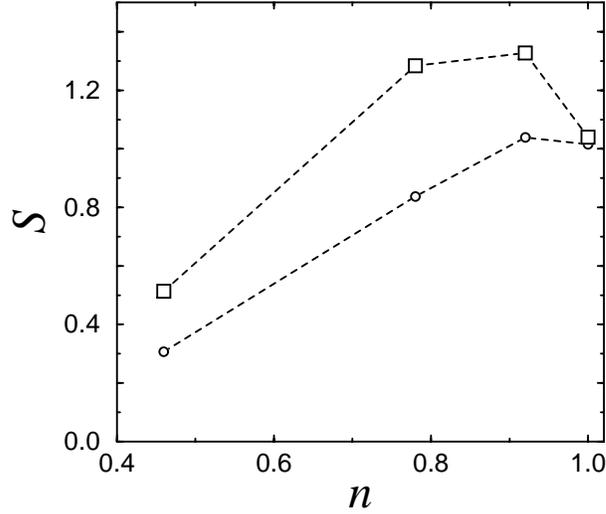}
\end{center}
\caption{
The integrated pairing correlation plotted against 
the band filling $n$ for a $10\times 10$ lattice with 
$U=0$ $(\circ)$ or $U=1$ (square).\cite{Kuroki}
}
\label{hubbardQMCndep}
\end{figure}

How about the band-filling ($n$) dependence?  
We have calculated the long-range 
part of the correlation, 
$S \equiv \sum_{r\geq 3}P(r)$, for various values of $n$ 
keeping $\Delta\varepsilon^0<0.01t$ throughout. 
The result, displayed in Fig.\ref{hubbardQMCndep}, 
shows that the enhancement in $S$ for $U=1$ has a maximum around 
a finite doping. 
Thus the message here is 
that the d$_{x^2-y^2}$ pairing is favoured {\it near, but 
not exactly at}, half-filling.  
The fact that the better nesting does not necesarrily 
imply the more enhanced pairing correlation has also 
been shown in another numerical work in the contex of 
an organic superconductivity.\cite{KAorganic}  

\section{Which is more favourable for superconductivity, 2D or 3D?}

The theoretical results described so far indicate that 
the superconductivity near the AF instability in 2D 
has a `low $T_C$' $\sim O(0.01t)$ ($t$: transfer integral), 
i.e., two orders of magnitude smaller 
than the original electronic energy, but still 
`high $T_C$' $\sim O(100$ K) for $t\sim O(1$ eV).  
So, identifying the conditions for higher $T_C$ in the repulsive 
Hubbard model is one 
of the most fascinating goals of theoretical studies.  
For instance, while the high-$T_C$ cuprates are layer-type 
materials with Cu$_2$ planes in which the supercurrent flows, 
the question is whether the two-dimensionality is promoting 
or degrading the superconductivity.

Hence Arita et al\cite{2d3d} have questioned:
(i) Is 2D system more favourable for spin-fluctuation
mediated superconductivity than in three dimensions(3D)?
(ii) Can other pairing, such as a triplet $p$-pairing in the presence of 
ferromagnetic spin fluctuations, become competitive?
We take the single-band, repulsive Hubbard model as a 
simplest possible model, and look into the pairing with 
the FLEX method in ordinary (i.e., square, trianglar, fcc, bcc, etc) 
lattices in 2D and 3D.  The FLEX method 
has an advantage that systems having large spin fluctuations can be handled.  

As for 3D systems, Scalapino {\it et al}\cite{Scalapino}
showed for the Hubbard model that paramagnon
exchange near a spin-density wave instability gives rise to
a strong singlet $d$-wave pairing interaction, 
but $T_C$ was not discussed there.
Nakamura {\it et al}\cite{Nakamura}
extended Moriya's spin fluctuation theory of superconductivity\cite{Moriya2} 
to 3D systems, 
and concluded that $T_C$ is similar between the 2D and 3D cases 
provided that common 
parameter values (scaled by the band width) are taken. 
However, the parameters there are phenomelogical ones, 
so we wish to see whether the result remains valid 
for microscopic models.

As for the triplet pairing, the possibility of triplet pairing mediated by 
ferromagnetic fluctuations has been investigated
for superfluid $^3{\rm He}$\cite{Leggett}, 
a heavy fermion system ${\rm UPt_3}$\cite{heavyFermion}, and
most recently, an oxide ${\rm Sr_2RuO_4}$\cite{Sr}. 
It was shown that ferromagnetic fluctuations favour triplet pairing 
first by Layzer and Fay\cite{Layzer}
before the experimental observation of p-wave pairing in $^3{\rm He}$.
For the electron gas model, Fay and Layzer\cite{Layzer} and later
Chubukov\cite{Chubukov} has 
extended the Kohn-Luttinger theorem\cite{KohnLutt}
to $p$-pairing for 2D and 3D electron gas in the dilute limit.
Takada\cite{Takada} discussed the possibility of $p$-wave superconductivity
in the dilute electron gas with the Kukkonen-Overhauser model\cite{KO}.
As for lattice systems, 2D Hubbard model with large enough 
next-nearest-neighbor hopping $(t')$ has been shown to exhibit 
$p$-pairing for small band fillings.\cite{ChubukovLu} 
Hlubina\cite{Hlubina99} reached a similar conclusion by evaluating 
the superconducting vertex in a perturbative way.\cite{Takahashi}
However, the energy scale of the $p$-pairing in the Hubbard model, 
i.e., $T_C$, has not been evaluated so far.

Here we show that 
(i) $d$-wave instability mediated by AF 
spin fluctuation in 2D square lattice
is much stronger than those in 3D, while 
(ii) $p$-wave instability
mediated by ferromagnetic spin fluctuations in 2D are much weaker than 
the $d$-instability.  
These results, which cannot be predicted a priori, 
suggest that for the Hubbard model 
the `best' situation for the pairing instability 
is the 2D case with dominant AF fluctuations.

We consider the single-band Hubbard model 
with the transfer energy $t_{ij}=t (=1$ hereafter) for nearest 
neighbors along with $t_{ij}=t'$ for second-nearest neighbors, 
which is included to incorporate the band structure dependence.
The FLEX starts from a set of skeleton diagrams 
for the Luttinger-Ward functional to generate
a ($k$-dependent) self energy 
based on the idea of Baym and Kadanoff\cite{Baym}. 
Hence the FLEX approximation is a self-consistent perturbation
approximation with respect to on-site interaction $U$.

To obtain $T_C$, we solve the eigenvalue ({\'E}liashberg) equation, 
\begin{eqnarray}
\lambda\Sigma^{(2)}(k)&=&\frac{T}{N}
\sum_{k'}
\Sigma^{(2)}(k')|G(k')|^2 V^{(2)}(k-k'),
\label{eliash}
\end{eqnarray}
where $\Sigma^{(2)}(k)$ is the anomalous self energy, 
$k\equiv (\Vec{k},i\omega_n)$ with $\omega_n=(2n-1)\pi T$ 
being Matsubara frequencies, and 
the pairing interaction, $V^{(2)}$, 
comprises contributions from the transverse spin fluctuations, 
longitudinal spin fluctuations 
and charge fluctuations, namely,
\begin{eqnarray*}
V^{(2)}(k,k')&=&-U^2\left[\frac{1}{2}\chi_{\rm ch}(k-k') \right.\\
&&\left. -\frac{1}{2}\chi^{zz}(k-k')+\chi^{\pm}(k+k')\right]\\
&=&
- \left[ \frac{U^3\chi_{\rm irr}^2(k-k')}{1-U^2\chi_{\rm irr}^2(k-k')} \right]
- \left[ \frac{U^2\chi_{\rm irr}(k+k')}{1-U\chi_{\rm irr}(k+k')} \right] 
\end{eqnarray*}
Here $\chi_{\rm ch}$ is the charge susceptibility, 
$\chi^{zz} (\chi^{\pm})$ 
the longitudinal (transverse) spin susceptibility, and 
\[
\chi_{\rm irr}(q)\equiv -(T/N)\sum_k G(k)G(k+q)
\]
the irreducible susceptibility constructed from the dressed Green's function.  
The dressed Green's function, $G(k)$, obeys the Dyson equation, 
\begin{equation}
{G(k)}^{-1} = {G^0(k)}^{-1}-\Sigma(k),
\label{Dyson}
\end{equation}
where $G^0$ is the bare Green's function, 
and $\Sigma$ the self energy with
\begin{eqnarray}
\label{Chi}
\Sigma(k)=\frac{1}{N}\sum_{q} G(k-q)V^{(1)}(q).
\end{eqnarray}
If we take RPA-type bubble and ladder diagrams for the 
interaction $V^{(1)}$, we have
\begin{eqnarray}
V^{(1)}(q)&=&\frac{1}{2}U^2\chi_{\rm irr}(q) 
\left[ \frac{1}{1+U\chi_{\rm irr}(q)} \right] \nonumber\\ 
&+&\frac{3}{2}U^2\chi_{\rm irr}(q) 
\left[ \frac{1}{1-U\chi_{\rm irr}(q)} \right] 
-U^2\chi_{\rm irr}(q), \nonumber
\end{eqnarray}
which completes the set of equations.

Since we have $\Sigma^{(2)}(k)=\Sigma^{(2)}(-k)$ for the spin-singlet pairing 
whereas $\Sigma^{(2)}(k)=-\Sigma^{(2)}(-k)$ for the spin-triplet pairing,
$V^{(2)}(k,k')$ becomes a function of $k-k'=q$ with
\begin{eqnarray}
V^{(2)}(q)=
-\frac{3}{2} \left[ \frac{U^2\chi_{\rm irr}(q)}{1-U\chi_{\rm irr}(q)} \right]
+\frac{1}{2} \left[ \frac{U^2\chi_{\rm irr}(q)}{1+U\chi_{\rm irr}(q)} \right]
\label{pair1}
\end{eqnarray}
for the singlet pairing, and
\begin{eqnarray}
V^{(2)}(q)=\frac{1}{2} \left[ \frac{U^2\chi_{\rm irr}(q)}{1-U\chi_{\rm irr}(q)} \right]
+\frac{1}{2} \left[ \frac{U^2\chi_{\rm irr}(q)}{1+U\chi_{\rm irr}(q)} \right] 
\label{pair2}
\end{eqnarray}
for the triplet pairing. 
$T=T_C$ is identified as the temperature at which 
the maximum eigenvalue $\lambda_{\rm Max}$ reaches unity.

\begin{figure}
\begin{center}
\includegraphics[width=1.0\textwidth]{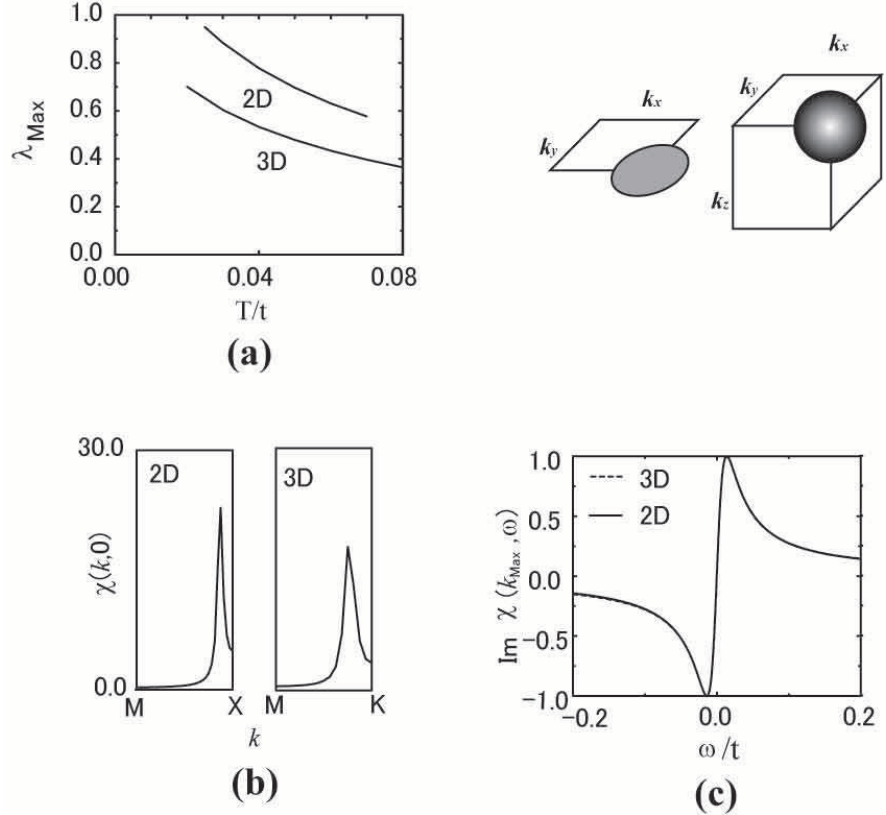}
\end{center}
\caption{
(a) The maximum eigenvalue of the {\'E}liashberg equation 
against temperature 
for the Hubbard model on a square lattice with 
$n=0.85$ and $U=4$ 
and on a cubic lattice with $n=0.8$, the second-neighbour 
hopping $t'=-0.2$ and $U=8$.  
$\chi(\Vec{k},0)$ (b) as a function of wavenumber 
and ${\rm Im}\chi(\Vec{k}_{\rm Max},\omega)$ (c) 
(normalised by its maximum) as a function of $\omega/t$ 
at $T=0.03t$ are also shown for the two lattices. 
Top right panel schematically shows the regions, 
in respective dimensions, that contribute to the pairing.
}
\label{2d3dresult}
\end{figure}

Let us start with the 2D case having strong AF fluctuations.  
We have first obtained 
$\chi_{\rm RPA}(q) = \chi_0 /(1-U\chi_0)$ as a function of the momentum 
for the Hubbard model on a nearly half-filled ($n=0.85)$ square lattice, where 
a dominant AF spin fluctuation is seen 
as $\chi_{\rm RPA}$ peaked around $(\pi,\pi)$.  
We can then plug this into the {\'E}liashberg equation (\ref{eliash}) to 
plot in Fig.\ref{2d3dresult}(a) 
$\lambda_{\rm Max}$ as a function of temperature $T$.  

$T_C$ is identified as the temperature at which 
$\lambda_{\rm Max}$ becomes unity, which occurs at $T\sim 0.02$ 
for the square lattice, in accord with previous results\cite{Dahm,finitetc}.

If we move on to
the case with ferromagnetic spin fluctuations where 
triplet pairing is expected, this situation can be realised 
for relatively large $t'(\simeq 0.5)$ 
and away from half-filling in the 2D Hubbard model.
Physically, the van Hove singularity shifts toward the band bottom 
with $t'$, and the large density of states at the Fermi level 
for the dilute case favours the ferromagnetism.  
We have found that $\lambda_{\rm Max}$ becomes largest for $n=0.3$, 
$t'=0.5$.
%

$\chi_{\rm RPA}$ is indeed peaked at $\Gamma$ ($\Vec{k}=(0,0)$).  
The question then is the behavior of 
$\lambda_{\rm Max}$ as a function of $T$, 
which shows that $\lambda_{\rm Max}$ is much smaller than that in the AF case.

A low $T_C$ for the ferromagnetic case 
contrasts with a naive expectation from the BCS picture, in which 
the Fermi level located around a peak in the density of states 
favours superconductivity.  
We may trace back two-fold reasons why this does not apply.  
First, if we look at the dominant ($\propto 1/[1-U\chi_0(q)]$) term of 
the pairing potential $V^{(2)}$ itself in eqs. (\ref{pair1}) and 
(\ref{pair2}), the triplet pairing interaction is only one-third of that
for singlet pairing.  
Second, the factor $|G|^2$ for the ferromagnetic case 
is smaller than that in the AF case, 
which implies that the self-energy correction is larger in the former.  
Larger self-energy (smaller $|G|^2$) works unfavourably 
for superconductivity as seen in 
the {\'E}liashberg equation (\ref{eliash}). 
When we take a larger repulsion $U$ to increase the 
triplet pairing attraction (susceptibility), 
this makes the self-energy correction 
even stronger.

Let us now move on to the case of $d$-wave pairing 
in the 3D Hubbard model, for which FLEX was first applied 
by Arita et al\cite{2d3d}. 
In simple-cubic systems, 
we find that the $\Gamma_{3}^{+}$ representation of O$_{h}$
group\cite{Sigrist}
has the largest $\lambda_{\rm Max}$.  
We have found that $\lambda_{\rm Max}$ for this symmetry
becomes largest for $n=0.8$, $t'=-0.2\sim -0.3$ and $U=8\sim 10$.
In Fig. \ref{2d3dresult}(a), we superpose $\lambda_{\rm Max}$
as a function of $T$, where 
we can immediately see that the pairing tendency in 3D is much {\it weaker} 
than that in 2D.  

Why is the $d$-superconductivity much stronger
in 2D than in 3D?  We can pinpoint the origin by looking at 
the various factors involved in the {\'E}liashberg equation. 
Namely we question the height of $V^{(2)}$ and $|G|^2$ 
along with the width of the region, both in the momentum sector 
and in the frequency sector, over which $V^{(2)}(k)$ 
contributes to the summation over $k\equiv (\Vec{k},i\omega_n)$.
We found that the maximum of $|G|^2$ 
is in fact larger in 3D than in 2D.  
The width of the peak in $\chi_{\rm RPA}$ on the frequency 
and momentum axes 
is surprisingly similar between 2D and 3D 
as displayed in Fig.\ref{2d3dresult}(b)(c).  
Note that if the frequency spread of the susceptibility 
scaled not with $t$ but with the {\it band width}, 
as Nakamura {\it et al}\cite{Nakamura}
have assumed, $\lambda_{\rm Max}$ would have become larger. 
Now, $\lambda$ in the {\'E}liashberg equation (\ref{eliash}) 
is $\propto (a/L)^D$, where 
$L$ is the linear dimension of the system and 
$a$ the width in the momentum space for the 
effective attraction, this factor is much smaller in 3D than 
in 2D as far as the main contribution of $V^{(2)}$ to the 
pairing occurs through special points in the k-space 
(e.g., $(\pi,\pi)$ or $(\pi,\pi,\pi)$ for the antiferromagnetic 
spin fluctuation exchange pairing).
So we can conclude that this is the main reason why 2D 
is more favourable than 3D.  

To summarise this section, 
$d$-pairing in 2D is the best situation for the repulsion 
originated (i.e., spin fluctuation mediated) superconductivity
in the Hubbard model.  
Monthoux and Lonzarich\cite{MonLon} have also concluded for 2D systems, 
by making use of a phenomenological approach, 
that the $d$-wave pairing is much stronger than $p$-wave pairing,
which is consistent with the present result.
In this sense, the layer-type cuprates do seem to hit upon the 
right situation.  

This is as far as one-band model 
having simple Fermi surfaces are concerned.  
Indeed, if we turn to heavy fermion superconductors, 
for instance, in which the pairing 
is thought to be meditated by spin fluctuations,
the $T_C$, when normalized by the band width $W$, 
is known to be of the order of $0.001W$.   
Since the present result indicates that $T_C$, normalized by $W$, 
is $\sim 0.0001W$ at best in the 3D Hubbard model, 
we may envisage that the heavy fermion system 
must exploit other factors such as the multiband.  
Neverthless, recent experimental finding\cite{rh} that a heavy-fermion 
compound Ce(Rh,Ir,Co)In$_5$ has the higher 
$T_c$ for the more two-dimensional lattice (with larger $c/a$) is 
consistent with our prediction.

\section{How to realise higher $T_C$ in anisotropic pairing --- 
disconnected Fermi surfaces}

Ironically, the main question about the superconductivity 
from the electronic mechanism is ``why is 
$T_C$ so low?", which has been repeatedly raised in literatures.  
Namely, one remarkable point is $T_c \sim O(0.01t)$, 
esitmated for the repulsive Hubbard model in the 
two-dimensional (2D) square lattice, is 
{\it two orders of magnitudes smaller than} the 
starting electronic energy (i.e., the hopping integral $t$), 
although this gives the right order for the curates' $T_c$. 
We have seen that even the best case, as far as these ordinary lattices are 
concerned, has $T_c \sim O(0.01t)$.  
As discussed in Ref. \cite{Kuroki&Arita}, 
there are good reasons 
why $T_c$ is so low: One reason is the effective 
attraction mediated by spin fluctuations is 
much weaker than the original electron-electron interaction, $U$. 
Another important reason is the 
presence of nodes in the superconducting gap function greatly 
reduces $T_c$:  While the main pair-scattering, 
across which the gap function has opposite signs to 
make the effective interaction attractive, 
there are other pair scatterings around the nodes 
that have negative contributions to the effective attraction 
by connecting $k$-points on which the gap has the same sign. 

So a next important avenue to explore is: 
can we improve the situation by going over to multiband systems.  
Kuroki and Arita \cite{Kuroki&Arita} 
have shown that this is indeed the case if we 
have {\it disconnected Fermi surfaces}.  In this case 
$T_c$ is dramatically enhanced, because the sign change 
in the gap function can avoid the Fermi pockets, 
where all the pair-scattering processes contribute 
positively. \cite{Kuroki&Arita,NTT} 
This has been numerically shown to be the case for 
the triangular lattice (for spin-triplet pairing) 
\cite{KAfortri} and 
a squre lattice with a period-doubling \cite{Kuroki&Arita}, 
where $T_c$ as estimated with FLEX is as high as $O(0.1t)$.  

\begin{figure}
\begin{center}
\includegraphics[width=1.0\textwidth]{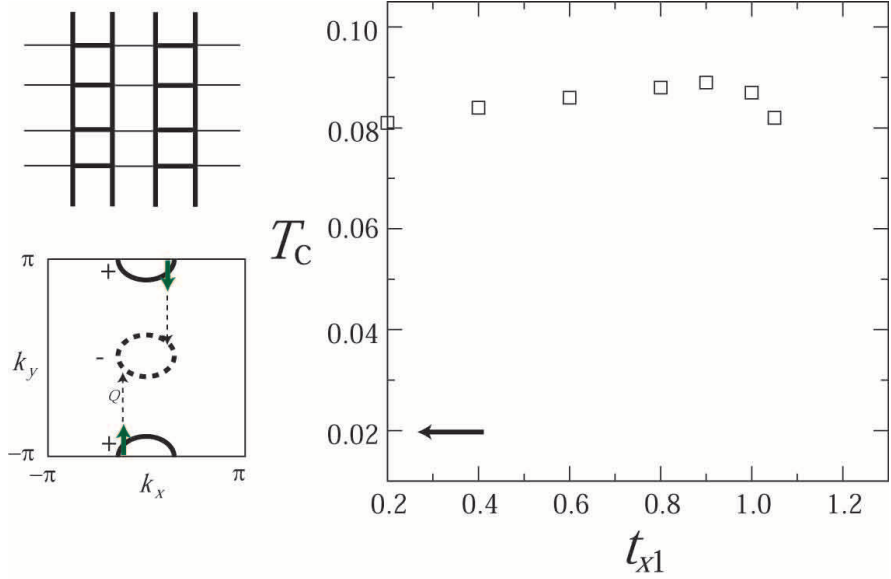}
\end{center}
\caption{
The $T_C$ estimated for the lattice depicted in the top left panel 
having disconnected Fermi surface (bottom left) as a function of 
the weak transfer (thin lines in the top left panel; 
$t_{x1}$).\cite{Kuroki&Arita}  
The arrow indicates typical $T_C$ for ordinary (e.g., square) 
lattices.
}
\label{discon}
\end{figure}

To be more precise, the key ingredients are: 
(a) when the Fermi surface is nested, 
the spin susceptibility $\chi(\Vec{q},\omega)$ 
has a peak. 
(b) When a multiband system with a disconnected Fermi surface 
has an inter-pocket nesting 
(i.e., strong inter-pocket pair scattering 
and weak intra-pocket one)
the gap function has the same sign ($s$-wave symmetry) 
within each pocket, and the nodal lines can happily 
run in between the pockets.   
The estimated $T_c$ for two-dimensional (2D) Hubbard model on such lattices is 
indeed almost an order of magnitude higher, $T_c\sim 0.1t$, 
as displayed in Fig.\ref{discon} along with the lattice structure.

\begin{figure}
\begin{center}
\includegraphics[width=1.0\textwidth]{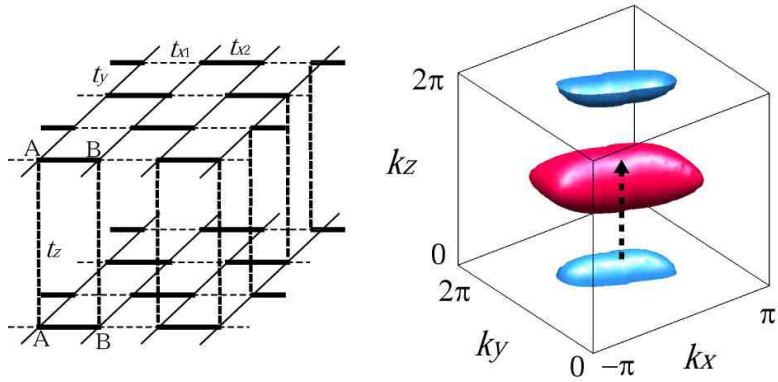}
\end{center}
\caption{
Interband nesting (arrow in the right panel) 
in 3D on a disconnected Fermi surface, 
which is exemplified here for the stacked bond-alternating 
lattice (left).
}
\label{3dinterband}
\end{figure}

As for the dimensionality of the system, 
we have shown above that 2D systems are generally more 
favourable than 3D systems 
as far as the spin-fluctuation-mediated superconductivity 
in ordinary lattices (square, triangular, fcc, bcc, etc) are concerned.  
Now, if one puts the idea for the disconnected Fermi 
surface on the above observation on the dimensionality, 
a natural question is:  
can we conceive 3D lattices having disconnected Fermi surfaces 
that have high $T_c$'s.  More specifically, 
can the disconnected Fermi surface overcome the 
disadvantage of 3D?  
If we express our idea more explicitly, what we have in mind 
is the {\it interband nesting} (or Suhl-Kondo process 
in its broader context) in the 3D disconnected Fermi 
surface as depicted in Fig.\ref{3dinterband}.  


There, the nesting vector runs across the two bands, and this 
is envisaged to give the attractive pair-scattering interaction.  
So the gap function should be nodeless within each band, while the 
gap has opposite signs between the two bands.  

In our most recent study\cite{onari3d} 
we have found that a stacked bond-alternating 
lattice (Fig.\ref{3dinterband}) has a compact and disconnected 
(i.e., a pair of ball-like) Fermi pockets.  
We have shown that $T_c$ is $O(0.01t)$, 
which is the same order of that for the square lattice, and remarkably 
high for a 3D system.
We have further found that the $T_c$ can be made even 
higher ($\sim O(0.1t))$ in a model in which the 
original Kuroki-Arita 2D system having 
disconnected Fermi surface is stacked.
So the final message obtained here, starting from 1D and 
ending up with 3D, is that 3D material with considerably high $T_c$ 
can be expected if we consider appropriate 
lattice structures. 

\section{Closing remarks}

So we have seen the electronic properties of electron 
systems with short-range repulsive interactions, starting 
from 1D up to 3D systems.  While the quasi-1D ladders already 
contain seeds for the d-wave pairing, anisotropic pairing 
has more degrees of freedom in 2D and 3D where the 
topology (e.g., disconnected Fermi surfaces) of the Fermi 
surface can greatly favour higher Tc.  
Finally it would be needless to stress that the electron correlation is 
such a fascinating subject that there are many open questions 
to be explored.   
Among them, a question we can ask is what would happen to 
the superconductivity from the repulsive electron interaction 
when disorder is introduced in the system.  Then we have a 
problem of dirty superconductors, i.e., 
an interplay of interaction and disorder.  For ladders there are 
some discussions on this.   For instance, Kimura et al.\cite{Kimuratransp} 
have looked at the dirty double wire (i.e., two-band Tomonaga-Luttinger 
system with impurities), and noted that in the phases where 
the pairing correlation is dominant, the Anderson localisation 
is absent despite the system being quasi-1D.  How this would be 
extended to higher dimensions is an interesting issue.

Acknowledgements --- 
First and foremost, I wish to thank Professor Bernard Kramer 
for many years of interactions and discussions, dating back to 
the advent period when 
the Anderson localisation just began to 
form a main stream in the condensed matter physics.  
For the physics on Tomonaga-Luttinger and 
electron-mechanism superconductivity I would like to thank 
Kazuhiko Kuroki for collaboration.  I also wish to 
acknowledge Ryotaro Arita, Takashi Kimura, Miko Eto, Michele 
Fabrizio and Seiichiro Onari for 
collaborations.  Some of the works described here 
have been supported by 
Grants-in-aids from the Japanese Ministry of Education, 
and I also thank Yasumasa Kanada for a
support in `Project for Vectorised Supercomputing' 
in the numerical works.  



\begin{thebibliography}{99}
\bibitem{KohnLutt}W. Kohn and J.M. Luttinger, Phys. Rev. Lett. {\bf 15}, 
524 (1965).

\bibitem{Layzer} 
D. Fay and A. Layzer: Phys. Rev. Lett. {\bf 20}, 187 (1968); 
A. Layzer and D. Fay: Int. J. Magn. {\bf 1}, 135 (1971).
\bibitem{Takadaswave} The pairing for more dilute electron gas 
is shown to be s-wave, where the dynamical interaction 
becomes negative (see, \protect\cite{Takada}).

\bibitem{Tomonaga} S. Tomonaga: Prog. Theoret. Phys. {\bf 5}, 544 (1950).
\bibitem{TomLutreview} For reviews of the Tomonaga-Luttinger 
theory, see J. S\'{o}lyom: Adv. Phys. {\bf 28}, 201 (1979); 
V.J. Emery in {\it Highly Conducting One-Dimensional 
Solids}, ed. by J.T. Devreese et al.(Plenum, New York, 1979), p.247. 
\bibitem{Mattis} D.C. Mattis: {\it The Theory of Magnetism I} (Springer-Verlag, Berlin, 1988).

\bibitem{weak1} A. Luther and V.J. Emery:
Phys. Rev. Lett. {\bf 33}, 589 (1974).
\bibitem{weak2}
P.A. Lee: Phys. Rev. Lett. {\bf 34}, 1247 (1975).
\bibitem{weak3}
C.M. Varma and A. Zawadowski: Phys. Rev. B {\bf 32}, 7399 (1985).
\bibitem{weak4}
K. Penc and J. S\'{o}lyom: Phys. Rev. B {\bf 41}, 704 (1990).
\bibitem{weak5}
A.M. Finkel'stein and A.I. Larkin: 
Phys. Rev. B {\bf 47}, 10461 (1993).

\bibitem{nagaosaogawa} N. Nagaosa and T. Ogawa: Solid 
State Commun. {\bf 88}, 295 (1993).

\bibitem{spinpolTL} 
T. Kimura, K. Kuroki, H. Aoki and M. Eto: 
Phys. Rev. B {\bf 49}, 16852 (1994); 
T. Kimura, K. Kuroki, and H. Aoki: 
{\it ibid} {\bf 53}, 9572 (1996).

\bibitem{sassetti} M. Sassetti, F. Napoli and B. Kramer: 
Phys. Rev. B {\bf 59}, 7297 (1999).

\bibitem{SchulzAF} H.J. Schulz: Phys. Rev. B {\bf 34}, 6372 (1986). 
\bibitem{Haldane} F.D.M. Haldane: Phys. Lett. A {\bf 93}, 464 (1993). 
\bibitem{Nishiyama} Y. Nishiyama, N. Hatano and M. Suzuki:
J. Phys. Soc. Jpn {\bf 64}, 1967 (1994).
\bibitem{Dagotto1} E. Dagotto, J. Riera, and D.J. Scalapino: Phys. Rev. B 
{\bf 45}, 5744 (1992).
\bibitem{Rice} T.M. Rice, S. Gopalan and M. Sigrist: 
Europhys. Lett. {\bf 23}, 445 (1993); Physica B {\bf 199 \& 200}, 378 (1994).

\bibitem{Anderson} P.W. Anderson: Science {\bf 235}, 1196 (1987).
\bibitem{Uehara}M. Uehara, T. Nagata, J. Akimitsu, H. Takahashi, 
N. M{\^o}ri and K. Kinoshita: 
J. Phys. Soc. Jpn. {\bf 65}, 2764 (1996).

\bibitem{Finkelstein} 
A.M. Finkel'stein and A.I. Larkin: Phys. Rev. B {\bf 47}, 10461 (1993).
\bibitem{Balents}L. Balents and M.P.A. Fisher: Phys. Rev. B {\bf 53}, 
12133 (1996).
\bibitem{Fabrizio} M. Fabrizio: Phys. Rev. B {\bf 48}, 15838 (1993).
\bibitem{Fab} M. Fabrizio, A. Parola and E. Tosatti:
Phys. Rev. B {\bf 46}, 3159 (1992).
\bibitem{Nagaosa}N. Nagaosa and M. Oshikawa: 
J. Phys. Soc. Jpn. {\bf 65}, 2241 (1996).
\bibitem{Schulz2}H.J. Schulz: Phys. Rev. B {\bf 53}, R2959 (1996).
\bibitem{Noack1}R.M. Noack, S.R. White and D.J. Scalapino: 
Phys. Rev. Lett. {\bf 73}, 882 (1994).
\bibitem{Noack3}R.M. Noack, S.R. White and D.J. Scalapino:
Physica C {\bf 270}, 281 (1996).
\bibitem{Dagotto} E. Dagotto, J. Riera and D.J. Scalapino: 
Phys. Rev. B {\bf 45}, 5744 (1992). 
\bibitem{Sigrist}M. Sigrist and K. Ueda: Rev. Mod. Phys. {\bf 63}, 239 (1991).
\bibitem{Poilblanc3} D. Poilblanc, H. Tsunetsugu and T.M. Rice:
Phys. Rev. B {\bf 50}, 6511 (1994).
\bibitem{Tsunetsugu} 
H. Tsunetsugu, M. Troyer and T.M. Rice:
Phys. Rev. B {\bf 49}, 16078 (1994); {\it ibid.} {\bf 51}, 16456 (1995).

\bibitem{Hayward} C.A. Hayward {\it et al.}: 
Phys. Rev. Lett. {\bf 75}, 926 (1995); 
C.A. Hayward and D. Poilblanc: Phys. Rev. B {\bf 53}, 
11721 (1996).  

\bibitem{Sano} K. Sano: J. Phys. Soc. Jpn. {\bf 65}, 1146 (1996).

\bibitem{KAmultiTL} Multiband Tomonaga-Luttinger model that includes 
pair transfer between two bands [K.A. Muttalib and V.J. Emery, 
Phys. Rev. Lett. {\bf 57}, 1370 (1986)] or 
charge-charge coupling between the chains 
[K. Kuroki and H. Aoki, Phys. Rev. Lett. {\bf 72}, 2947 (1994)] 
has been studied. 

\bibitem{Suhl} H. Suhl, B.T. Mattis and L.R. Walker:
Phys. Rev. Lett. {\bf 3}, 552 (1959).
\bibitem{Kondo} J. Kondo:
Prog. Theor. Phys. {\bf 29}, 1 (1963).
\bibitem{Hirsch} J.E. Hirsch: Phys. Rev. B {\bf 31}, 4403 (1985).
\bibitem{MC1} M. Imada and Y. Hatsugai: 
J. Phys. Soc. Jpn. {\bf 58}, 3572 (1989).
\bibitem{MC2}
N. Furukawa and M. Imada: J. Phys. Soc. Jpn. {\bf 61}, 3331 (1992).
\bibitem{MC3}
S. Sorella, E. Tosatti, S. Baroni, R. Car and 
M. Parrinello:   
Int. J. Mod. Phys. B {\bf 1}, 993 (1988).
\bibitem{MC4}
S.R. White, D.J. Scalapino, R.L. Sugar, E.Y. Loh, 
J.E. Gubernatis and R.T. Scalettar:  
Phys. Rev. B {\bf 40}, 506 (1991).
\bibitem{MC5}
W. von der Linden, I. Morenstern and H. de Raedt: 
Phys. Rev. B {\bf 41}, 4669 (1990).

\bibitem{Noack2} R.M. Noack, S.R. White and D.J. Scalapino:
Europhys. Lett. {\bf 30}, 163 (1995); Physica C {\bf 270}, 281 (1996).

\bibitem{Asai} Y. Asai: Phys. Rev. B {\bf 52}, 10390 (1995).
\bibitem{Yamaji} 
K. Yamaji and Y. Shimoi: Physica C {\bf 222}, 349 (1994); 
K. Yamaji, Y. Shimoi, and T. Yanagisawa: 
{\it ibid} {\bf 235-240}, 2221 (1994).

\bibitem{Kuroki}K. Kuroki and H. Aoki: Phys. Rev. B {\bf 56}, R14287 (1997);
J. Phys. Soc. Jpn {\bf 67}, 1533 (1998).
\bibitem{pmc} 
S. Sorella {\it et al.}: Int. J. Mod. Phys. B {\bf 1},
993 (1988); S.R. White {\it et al.}: Phys. Rev. B {\bf 40},
506 (1989); M. Imada and Y. Hatsugai: J. Phys. Soc. Jpn. {\bf 58},
3752 (1989).
\bibitem{tJcom} 
Three-leg $t$-$J$ ladder has also been examined by 
T.M. Rice, S. Haas, M. Sigrist and F.C. Zhang: 
Phys. Rev. B {\bf 56}, 14655 (1997) with exact diagonalisation 
and mean-field approximation, and 
by 
S.R. White and D.J. Scalapino: Phys. Rev. B {\bf 57}, 3031 (1998) 
with DMRG.

\bibitem{Azuma} M. Azuma, Z. Hiroi, M. Takano, K. Ishida and Y. Kitaoka: 
Phys. Rev. Lett. {\bf 73}, 3463 (1994).
\bibitem{Ishida1} K. Ishida, Y. Kitaoka, Y. Tokunaga, S. Matsumoto, 
K. Asayama, M. Azuma, Z. Hiroi and M. Takano:
Phys. Rev. B {\bf 53}, 2827 (1996). 
\bibitem{Kojima} K. Kojima, A. Karen, G.M. Luke, B. Nachumi, W.D. Wu, 
Y.J. Uemura, M. Azuma and M. Takano:  
Phys. Rev. Lett. {\bf 74}, 2812 (1995). 
\bibitem{Hiroi2} Z. Hiroi and M. Takano:
Nature {\bf 377}, 41 (1995). 
\bibitem{Mayaffre} H. Mayaffre, P. Auban-Senzier, D. J{\'e}rome, D. Poilblanc,
C. Bourbonnais, U. Ammerahl, G. Dhalenne, and A. Revcolevschi:
Science {\bf 279}, 345 (1998).

\bibitem{Takashi1}T. Kimura, K. Kuroki and H. Aoki: 
Phys. Rev. B {\bf 54}, R9608 (1996).
\bibitem{Takashi2} T. Kimura, K. Kuroki and H. Aoki: 
J. Phys. Soc. Jpn. {\bf 66}, 1599 (1997); {\bf 67}, 1377 (1998).

\bibitem{Arrigoni}E. Arrigoni: Phys. Lett. A {\bf 215}, 91 (1996); 
Phys. Rev. Lett. {\bf 83} 128 (1999); Phys. Rev. B {\bf 61} 7909 (2000).
See also, for the even-odd conjecture for the spin gap, 
U. Ledermann, K. Le Hur, and T.M. Rice, Phys. Rev. B {\bf 62}, 16383 (2000); 
U. Ledermann, Phys. Rev. B {\bf 64}, 235102 (2001).

\bibitem{Schulz3} 
H.J. Schulz in {\it Correlated Fermions and Transport in Mesoscopic Systems}, 
ed. by T. Martin, G. Montambaux and J.T.T. Van 
(Editions Frontieres, Gif-sur-Yvette, 1996), p. 81.

\bibitem{dpKA} To be precise the CuO$_2$ plane consists of a 
square array of Cu d orbitals and O p orbitals, which are usually 
modelled by the ``d-p" model.  Quantum Monte Carlo has indeed 
detected a long-ranged pairing correlation [K. Kuroki and 
H. Aoki: Phys. Rev. Lett. {\bf 76}, 4400 (1996)].  
The d-p model can be mapped to a one-band Hubbard model 
[see M.S. Hybertsen, E.B. Stechel, M. Schl\"{u}ter, and 
D.R. Jennison: Phys. Rev. B {\bf 41}, 11068 (1990); 
M.S. Hybertsen and M. Schl\"{u}ter in 
{\it New Horizons in Low-Dimensional 
Electron Systems} ed. by H. Aoki et al (Kluwer, Dordrecht, 1992) p.229, 
and refs therein].
\bibitem{Andersonreview} P.W. Anderson: {\it A Carrier in Theoretical 
Physics} (World Scientific, Singapore, 1994) and refs therein.
\bibitem{Moriya1}T. Moriya, Y. Takahashi, and K. Ueda:
J. Phys. Soc. Jpn, {\bf 59}, 2905 (1990); 
Physica C {\bf 185-189}, 114 (1991).
\bibitem{Moriya2}T. Ueda, T. Moriya, and Y. Takahashi in 
{\it Electronic Properties and Mechanisms of High-$T_C$ Superconductors} 
ed. T. Oguchi {\it et al.} (North Holland, Amsterdam, 1992), p. 145; 
J. Phys. Chem. Solids {\bf 53}, 1515 (1992).
\bibitem{Moriya3}T. Moriya and K. Ueda: J. Phys. Soc. Jpn. {\bf 63}, 
1871, (1994).
\bibitem{Pines1}P. Monthoux, A. V. Balatsky, and D. Pines:
Phys. Rev. B {\bf 46}, 14803 (1992); 
\bibitem{Bickers} N.E. Bickers, D.J. Scalapino, and S.R. White: 
Phys. Rev. Lett. {\bf 62}, 961 (1989).
\bibitem{Dzyalo} I.E. Dzyaloshinskii: Zh.Eksp. Teor. Fiz. {\bf 93}, 
2267 (1987).
\bibitem{Schulz} H.J. Schulz: Europhys. Lett. {\bf 4}, 609 (1987).
\bibitem{ScalapinoPhysRep} D.J. Scalapino: Phys. Rep. {\bf 250}, 
329 (1995).
\bibitem{Alvarez}J.V. Alvarez, J. Gonzalez, F. Guinea and M.A.H. Vozmediano: 
J Phys. Soc. Jpn {\bf 67}, 1868 (1998).
\bibitem{FLEX}N. E. Bickers, D. J. Scalapino, and S. R. White:
Phys. Rev. Lett. {\bf 62}, 961 (1989); 
N. E. Bickers and D. J. Scalapino: Ann. Phys. (N. Y.)
{\bf 193}, 206 (1989).
\bibitem{Dahm}T. Dahm and L. Tewordt: Phys. Rev. B {\bf 52}, 1297 (1995).
\bibitem{Deisz}J.J. Deisz, D. W. Hess, and J. W. Serene:
Phys. Rev. Lett. {\bf 76}, 1312 (1996).
\bibitem{DMST} E. Dagotto {\it et al.}: Phys. Rev. B {\bf 41}, 811 (1990).
\bibitem{FOP} G. Fano, F. Ortolani, and A. Parola: Phys. Rev. B {\bf 42},
6877 (1990).
\bibitem{White} S.R. White {\it et al.}: Phys. Rev. B {\bf 39}, 839 (1989); 
Phys. Rev. B {\bf 40}, 506 (1989).
\bibitem{Husslein} T. Husslein {\it et al.}: Phys. Rev. B {\bf 54}, 16179 
(1996).
\bibitem{Zhang} S. Zhang, J. Carlson, and J.E. Gubernatis: Phys. Rev. Lett.
{\bf 78}, 4486 (1997).
\bibitem{Giamarchi} T. Giamarchi and C. Lhuillier: Phys. Rev. B {\bf 43}, 
12943 (1991).
\bibitem{YamajiVMC} T. Nakanishi, K. Yamaji, and T. Yanagisawa: J. Phys. Soc.
Jpn. {\bf 66} 294 (1997).
\bibitem{Furukawa} N. Furukawa and M. Imada: J. Phys. Soc. Jpn.
{\bf 61}, 3331 (1992).
\bibitem{Moreo} A. Moreo: Phys. Rev. B {\bf 45}, 5059 (1992).
\bibitem{stab} G. Sugiyama and S.E. Koonin: Ann. Phys. {\bf 168},
1 (1986); S. Sorella {\it et al.}: Int. J. Mod. Phys. B {\bf 1},
993 (1988); S.R. White {\it et al.}: Phys. Rev. B {\bf 40},
506 (1989); M. Imada and Y. Hatsugai: J. Phys. Soc. Jpn., {\bf 58},
3752 (1989).
\bibitem{KTA} K. Kuroki, T. Kimura, and H. Aoki: Phys. Rev. B {\bf 54},
R15641 (1996).
\bibitem{LeeRead} P.A. Lee and N. Read: Phys. Rev. Lett. {\bf 58}, 2691
(1987).
\bibitem{Newns} D.M. Newns {\it et al.}: Phys. Rev. Lett. {\bf 69}, 
1264 (1992).
\bibitem{Mark} R.S. Markiewicz {\it et al.}: Physica C {\bf 217}, 381 (1993)
and references therein.
\bibitem{Gonzalez} J. Gonzalez and J.V. Alvarez:
Phys. Rev. B {\bf 56}, 367 (1997).


\bibitem{KAorganic} K. Kuroki and H. Aoki: 
Phys. Rev. B {\bf 60}, 3060 (1999).

\bibitem{2d3d} R. Arita, K. Kuroki and H. Aoki: 
Phys. Rev. B {\bf 60}, 14585 (1999); 
J. Phys. Soc. Jpn. {\bf 69}, 1181 (2000).
\bibitem{Scalapino}D. J. Scalapino, E. Loh, Jr., and 
J. E. Hirsh: Phys. Rev. B {\bf 34}, 8190 (1986).
\bibitem{Nakamura}S. Nakamura, T. Moriya and K. Ueda:
J. Phys. Soc. Jpn {\bf 65}, 4026 (1996).
\bibitem{Leggett}A. J. Leggett: Rev. Mod. Phys. {\bf 47}, 331 (1975).
\bibitem{heavyFermion}H. Tou {\it et al.}: 
Phys. Rev. Lett. {\bf 77}, 1374 (1996);
{\it ibid}, {\bf 80}, 3129 (1998).
\bibitem{Sr}Y. Maeno {\it et al.}: 
Nature {\bf 372}, 532 (1994); 
T. M. Rice, M. Sigrist: J. Phys. Condens. Matter {\bf 7}, L643 (1995).
\bibitem{Chubukov}
M.Y. Kagan and A.V. Chubukov: JETP Lett. {\bf 47}, 614 (1988); 
A.V. Chubukov: Phys. Rev. B {\bf 48}, 1097 (1993).

\bibitem{Takada}Y. Takada: Phys. Rev. B {\bf 47}, 5202 (1993).
\bibitem{KO}C. A. Kukkonen and A. W. Overhauser: 
Phys. Rev. B {\bf 20}, 550 (1979).
\bibitem{ChubukovLu}A. V. Chubukov and J. P. Lu: Phys. Rev. B {\bf 46}, 
11163 (1992).
\bibitem{Hlubina99}R. Hlubina: Phys. Rev. B {\bf 59}, 9600 (1999).
\bibitem{Takahashi} 
H. Takahashi [J. Phys. Soc. Jpn, {\bf 68}, 194 (1999)] 
on the other hand concludes that $p$-wave channel is most attractive 
for dilute ($n\sim 0.1$) 2D Hubbard model on the $t'=0$ square lattice.
\bibitem{Baym}G. Baym and L.P. Kadanoff: Phys. Rev. {\bf 124}, 
287 (1961); G. Baym: Phys. Rev. {\bf 127}, 1391 (1962).
\bibitem{finitetc} Here 
a finite $T_C$ in 2D systems is thought of as a measure of $T_C$ 
when the layers are stacked to Josephson-couple.  
\bibitem{MonLon}P. Monthoux and G. G. Lonzarich: 
Phys. Rev. B {\bf 59}, 14598 (1999).
\bibitem{rh}P.G. Pagliuso, R. Movshovich, A.D. Bianchi, M. Nicklas,
 N.O. Moreno, J.D. Thompson, M.F. Hundley, J.L. Sarrao, Z. Fisk: Physica
 B {\bf 312}, 129 (2002).
\bibitem{Kuroki&Arita} K. Kuroki, and R. Arita: Phys. Rev. B {\bf 64},
 024501 (2001).
\bibitem{NTT} T. Kimura, H. Tamura, K. Kuroki, K. Shiraishi, H. Takayanagi,
and R. Arita: Phys. Rev. B {\bf 68}, 132508 (2002).
\bibitem{KAfortri} K. Kuroki and R. Arita: Phys. Rev. B {\bf 63}, 174507 (2001).
\bibitem{onari3d} S. Onari, K. Kuroki, R. Arita and H. Aoki: 
Phys. Rev. B {\bf 65}, 184525 (2002); in preparation.
\bibitem{Kimuratransp}T. Kimura, K. Kuroki, and H. Aoki: 
Phys. Rev. B {\bf 51}, 13860 (1995).

\end{thebibliography}
\end{document}